# TIDAL FORCES IN COLD BLACK HOLE SPACETIMES

K. K.Nandi[1], A.Bhadra[2], P.M. Alsing[3], T.B.Nayak[1]

1. Department of Mathematics, University of North Bengal, Darjeeling (WB) 734430, India.
   e-mail: kamalnandi@hotmail.com

2. High Energy and Cosmic Ray Research Center, University of North Bengal, Darjeeling (WB) 734430, India.
   e-mail: bhadra@nbu.ernet.in

3. Albuquerque High Performance Computing Center, University of New Mexico, 1601 Central NE, Albuquerque, NM 87131, USA
   e-mail: alsing@ahpcc.unm.edu

**Abstract**

We investigate here the behavior of a few spherically symmetric static acclaimed black hole solutions in respect of tidal forces in the geodesic frame. It turns out that the forces diverge on the horizon of cold black holes (CBH) while for ordinary ones, they do not. It is pointed out that Kruskal-like extensions do not render the CBH metrics nonsingular. We present a CBH that is available in the Brans-Dicke theory for which the tidal forces do not diverge on the horizon and in that sense it is a better one.

## 1. Introduction

In vacuum Einstein's General Relativity (EGR), it is well known that the only static, spherically symmetric (SSS) black hole solutions are those given by the charged Reissner-Nordstrom (RN) and uncharged Schwarzschild ones. In the nonvacuum EGR, several black hole solutions are known. For instance: (i) The one discovered by Bekenstein [1], which is a solution of the Einstein conformally coupled scalar field theory. (ii) The dilaton-Maxwell solution discovered by Garfinkle, Horowitz and Strominger [2]. (iii) The solution for a nonlinear electromagnetic source found recently by Ayon-Beato and Garcia [3]. In the non-Einsteinian theories of gravity, too, some acclaimed black hole solutions exist in the literature. These theories include Brans-Dicke (BD) scalar-tensor theory [4] and the Weyl Integrable theory [5]. In the former, we have black hole solutions discussed by Campanelli and Lousto [6], Bronnikov et al [7] and in the latter, we have the Salim-Sautu [8] solutions. The list however is by no means exhaustive. The considered non-EGR family of solutions has infinite horizon area and zero Hawking temperature. Hence, such solutions have been baptized as cold black holes (CBH) by Bronnikov et al [7].

Let us now recall a very fundamental criterion which a given solution must satisfy in order that it can represent a black hole spacetime: The curvature tensor components, computed in the observer's (static or freely falling) orthonormal frame must be finite everywhere including the horizon [9]. This condition comes from the physical requirement that the tidal forces do not crush or tear apart an extended observer falling freely through the horizon.

We wish to examine in this paper how many of the above solutions satisfy the condition of finiteness of tidal forces near the horizon. It turns out that all the considered solutions except the CBH satisfy this criterion. We shall then examine a particular solution in the BD theory which also turns out to be a CBH but in which the malady of infinite tidal forces does not appear. In that sense, the solution merits as a better CBH than the available ones.

In Sec.2, the general expression for tidal forces in a freely falling frame is laid down. In Sections 3 and 4, different solutions in EGR and non-EGR respectively are tested. It is pointed out in Sec.5 that the Kruskal-like extensions do not render the CBH metrics nonsingular. Finally, in Sec.6, a better CBH in BD theory is discussed.

## 2. Tidal forces in a geodesic frame

Following the notations of Horowitz and Ross [10], consider the the general form of a SSS metric:

$$ds^2 = -\frac{F(r)}{G(r)}dt^2 + \frac{dr^2}{F(r)} + R^2(r)d\Omega^2, d\Omega^2 \equiv d\theta^2 + \sin^2\theta d\varphi^2. \quad (1)$$

In a static observer's orthonormal basis, the only nonvanishing components of the curvature tensor are $R_{0101}$, $R_{0202}$, $R_{0303}$, $R_{1212}$, $R_{1313}$ and $R_{2323}$. Radially freely falling observers with conserved energy E are connected to the static orthonormal frame by a local Lorentz boost with an instantaneous velocity given by

$$v = \left[1 - \frac{FE^{-2}}{G}\right]^{1/2}. \quad (2)$$

Then the nonvanishing curvature components in the Lorentz-boosted frame ($\wedge$) are [10]:

$$R_{\hat{0}\hat{1}\hat{0}\hat{1}} = R_{0101}, \quad (3)$$

$$R_{\hat{0}\hat{k}\hat{0}\hat{k}} = R_{0k0k} + \sinh^2\alpha(R_{0k0k} + R_{1k1k}), \quad (4)$$

$$R_{\hat{0}\hat{k}\hat{1}\hat{k}} = \cosh\alpha \sinh\alpha(R_{0k0k} + R_{1k1k}), \quad (5)$$

$$R_{\hat{1}\hat{k}\hat{1}\hat{k}} = R_{1k1k} + \sinh^2\alpha(R_{0k0k} + R_{1k1k}), \quad (6)$$

where k, l =2,3 and $\sinh\alpha = \frac{v}{\sqrt{1-v^2}}$. The tidal acceleration between two parts of the traveler's body is given by [11]:

$$\Delta a_{\hat{j}} = -R_{\hat{0}\hat{j}\hat{0}\hat{p}}\xi^{\hat{p}}, \quad (7)$$

where $\vec{\xi}$ is the vector separation between two parts of the body. All that we have to do now is to calculate the components in eqs.(3)-(6) for a given metric. If *any* of the components diverges as the horizon is approached, we say that the tidal forces physically disrupt the falling observer. Our strategy then is to first compute any one, say, $R_{\hat{0}\hat{2}\hat{0}\hat{2}}$, using eq.(4). If it is well behaved, then proceed to check if the same behavior is obtained for the rest of the components. If the answer is positive, we say that the solution represents an ordinary black hole. If $R_{\hat{0}\hat{2}\hat{0}\hat{2}}$ itself is not well behaved, we check no further and conclude that the solution might at best represent a cold black hole.

From the metric (1), we can rewrite eq.(4) as (k=2):

$$R_{\hat{0}\hat{2}\hat{0}\hat{2}} = -\frac{1}{R}\left[R''(E^2 G - F) + \frac{R'}{2}(E^2 G' - F')\right], \quad (8)$$

where primes on the right denote derivatives with respect to r. Now note that the conserved energy E can be decomposed as

$$E^2 = (F/G) + v^2(1-v^2)^{-1}(F/G) = E_s^2 + E_{ex}^2. \quad (9)$$

The first term represents the value of $E^2$ in the static frame ($E_s^2$) and the second term represents the enhancement in $E_s^2$ due to geodesic motion. Incorporating this, we can decompose $R_{\hat{0}\hat{2}\hat{0}\hat{2}}$ as follows:

$$R_{\hat{0}\hat{2}\hat{0}\hat{2}} = -\frac{1}{R}\left[\frac{R'}{2}\left(E_s^2 G' - F'\right)\right] - \frac{1}{R}\left(R''G + \frac{R'G'}{2}\right)E_{ex}^2$$
$$= R^{(s)}_{0202} + R^{(ex)}_{0202}. \tag{10}$$

It is easy to verify that the term $|R^{(s)}_{0202}|$ actually represents the curvature component in the static frame, viz., $R^{(s)}_{0202} = R_{0202}$. Thus, only the term $R^{(ex)}_{0202} (\equiv \sinh^2 \alpha (R_{0202} + R_{1212}))$ represents overall enhancement in curvature in the Lorentz-boosted frame over the static frame. It is this part that needs to be particularly examined as the observer approaches the horizon. Note also that the energy $E^2$ is finite (it can be normalized to unity) and so are $E_s^2$ and $E_{ex}^2$. As the horizon is approached, (F/G) $\to 0$, $v \to 1$ such that $E^2 \to E_{ex}^2$. Let us now proceed to test a few solutions.

### 3. EGR Solutions (Planck units)

*(a) RN solution*:

The metric is given by
$$ds^2 = -\left(1 - \frac{2m}{r} + \frac{Q^2}{r^2}\right)dt^2 + \left(1 - \frac{2m}{r} + \frac{Q^2}{r^2}\right)^{-1} dr^2 + r^2 d\Omega^2, \tag{11}$$

where Q is the electric charge. It can be readily verified that, due to a remarkable cancellation, $R^{(ex)}_{0202}$ is identically zero. In fact, since $R^{(ex)}_{0k0k}$ is the same as $\sinh^2 \alpha (R_{0k0k} + R_{1k1k})$, one can say that all the tensor components remain invariant under the Lorentz-boost. This is a peculiar feature of the RN geometry and is also shared by usual Schwarzschild geometry which is obtained by merely putting Q=0. All components in the static frame remain finite as r→2m. For example,

$$R_{0101} = -\frac{2m}{r^3} + \frac{3e^2}{r^4} \tag{12}$$

and so on. Hence it is concluded that the tidal forces do not diverge near the horizon either for the static or for moving observers.

*(b) Bekenstein solution:*

The metric is given by
$$ds^2 = -\left(1 - \frac{m}{r}\right)^2 dt^2 + \left(1 - \frac{m}{r}\right)^{-2} dr^2 + r^2 d\Omega^2, \tag{13}$$
$$m^2 = Q^2 + \tfrac{1}{3}q^2, \quad F_{\mu\nu} = Qr^{-2}\left(\delta_\mu^r \delta_\nu^t - \delta_\mu^t \delta_\nu^r\right),$$

and the conformal scalar field is given by $\phi = \dfrac{q}{r-m}$. As far as the metric is concerned, it has the extreme RN form and the same conclusions as above apply.

*(c) Garfinkle-Horowitz-Strominger (GHS) solutions:*

These are solutions to the low-energy string theory representing SSS charged black holes. The action is given by

$$S = \int d^4x \sqrt{-g}\left[-R + 2(\nabla\phi)^2 + e^{-2\phi}F^2\right] \qquad (14)$$

where $\phi$ and $F_{\mu\nu}$ are dilatonic and Maxwell fields respectively. A class of solutions is given by

$$ds^2 = -\left(1-\dfrac{2m}{r}\right)dt^2 + \left(1-\dfrac{2m}{r}\right)^{-1}dr^2 + r\left(r - \dfrac{Q^2 e^{-2\phi_0}}{mr}\right)d\Omega^2, \qquad (15)$$

$$e^{-2\phi} = e^{-2\phi_0}\left[1 - \dfrac{Q^2 e^{-2\phi_0}}{mr}\right], \quad F = Q\sin\theta d\theta \wedge d\varphi. \qquad (16)$$

It follows from eq.(4) that $R^{(ex)}_{0202}$ is simply proportional to $R''/R$ which, at the horizon $r_h=2m$, is

$$\dfrac{R''}{R} = -\dfrac{1}{2m(m+D)}, \qquad (17)$$

where D is the dilaton charge given by $D = -\dfrac{Q^2 e^{-2\phi_0}}{2m}$. If $Q^2 = 2m^2 e^{2\phi_0}$, then $\dfrac{R''}{R}\Big|_{r_h} \to \infty$, indicating that the tidal forces diverge at r=$r_h$. That is quite consistent with the conclusion of GHS that this value of $Q^2$ actually represents a transition between black holes and naked singularities. For $Q^2 \neq 2m^2 e^{2\phi_0}$, the solution (15) does represent a black hole. In the string frame, the metric is obtained by a conformal transformation $e^{2\phi}g_{\mu\nu}$, and it has been shown that, for $Q^2 < 2m^2 e^{2\phi_0}$, the new metric also represents a black hole [2]. According to our criterion, $\dfrac{R''}{R}\Big|_{r_h} \to finite$, which can be readily verified.

(d) *Black hole solution for nonlinear source:*

Recently, Ayon-Beato and Garcia [3] have proposed an exact SSS solution of EGR when the source is a nonlinear electrodynamic field. The resulting metric is

$$ds^2 = -\left[1 - \frac{2mr^2 e^{-\frac{q^2}{2mr}}}{(r^2+q^2)^{3/2}}\right]dt^2 + \left[1 - \frac{2mr^2 e^{-\frac{q^2}{2mr}}}{(r^2+q^2)^{3/2}}\right]^{-1} dr^2 + r^2 d\Omega^2 \quad (18)$$

where q is interpreted as the electric charge as the electric field expands asymptotically as

$$E = \frac{q}{r^2} + O(r^{-3}). \quad (19)$$

All curvature invariants are bounded everywhere *including* the origin. Evidently, $R_{0202}^{(ex)}$ is zero identically, indicating that the corresponding component of the tidal force is bounded too. In fact, it can be verified that all other components are also bounded. However, an undesirable feature of the solution, in our opinion, is that the horizon can not be *precisely* located in the spacetime as $g_{00}=0$ does not have an exact solution.

## 4. Non-EGR Solutions (Planck units)

*(e) Brans-Dicke Black Holes:*

Campanelli and Lousto [6] have shown that the BD theory admits a black hole solution which is different from the Schwarzschild one, the metric being given by

$$ds^2 = -\left(1 - \frac{2m}{r}\right)^{\alpha+1} dt^2 + \left(1 - \frac{2m}{r}\right)^{\beta-1} dr^2 + \left(1 - \frac{2m}{r}\right)^{\beta} r^2 d\Omega^2, \quad (20)$$

$$\phi(r) = \phi_0 \left(1 - \frac{2m}{r}\right)^{-\frac{\alpha+\beta}{2}}, \quad (21)$$

the coupling constant being given by

$$\varpi = -\frac{2(\alpha^2 + \beta^2 + \alpha\beta + \alpha - \beta)}{(\alpha+\beta)^2}. \quad (22)$$

According to Ref.[6], the solution represents a regular black hole for $\beta \leq -1$ and $\alpha-\beta+1 > 0$. In addition, if one requires that the metric should coincide with the PPN expansion, then we need to take $\alpha+\beta \to 0$, which implies $\varpi \to -\infty$. In this case, the solution takes on the form

$$ds^2 = -\left(1-\frac{2m}{r}\right)^{1-\beta} dt^2 + \left(1-\frac{2m}{r}\right)^{\beta-1} dr^2 + \left(1-\frac{2m}{r}\right)^{\beta} r^2 d\Omega^2, \quad (23)$$

and $\phi(r) = \phi_0 = const$. The parameter β plays the role of BD scalar hair. For this metric, all curvature invariants are finite, for β≤-1, as r→2m. However, it turns out that

$$R_{\hat{0}\hat{2}\hat{0}\hat{2}} = \frac{(1-\beta)m}{2r^3}\left[1+\frac{(\beta-2)m}{r}\right]^{-\beta-1} - \frac{2\beta(\beta-2)m^2}{r^4}\left(1-\frac{2m}{r}\right)^{-2} E_{ex}^2 \quad (24)$$

$$= R_{0202}^{(s)} + R_{0202}^{(ex)}.$$

We see that $R_{0202}^{(ex)} \to \infty$ as r→2m. Thus the horizon is singular. However, one might still argue that the value β=2 removes the divergence. But then the scalar invariant

$$I = R_{\alpha\beta\gamma\delta}R^{\alpha\beta\gamma\delta} = O[(r-2m)^{-2-2\beta}] \to \infty, \quad (25)$$

as r→2m. Black hole solutions (type B1), called CBHs, proposed by Bronnikov, Clement, Fabris and Constaninidis [7] also exhibit similar properties.

*(f) Black holes in Weyl Integrable Space Time (WIST):*

The spacetime described by Weyl integrable geometry follows from the action

$$S = \int \left(R + \xi\omega^{\alpha}_{;\alpha} - \tfrac{1}{2}e^{-2\omega}F_{\alpha\beta}F^{\alpha\beta} + e^{2\omega}g^{\mu\nu}\phi_{,\mu}\phi_{,\nu}\right)\sqrt{-g}d^4x, \quad (26)$$

where R is the scalar associated with the Weyl geometry, ω is the geometric scalar field and φ is an external scalar field. Salim and Sautu [8] proposed three classes of solutions. Let us consider only one of them:

$$ds^2 = -\left(1-\frac{\eta}{r}\right)^{m/\eta} dt^2 + \left(1-\frac{\eta}{r}\right)^{-m/\eta} dr^2 + \left(1-\frac{\eta}{r}\right)^{2m/\eta} r^2 d\Omega^2, \quad (27)$$

$$e^{\omega(r)} = e^{\omega_0}\left(1-\frac{\eta}{r}\right)^{-\sigma/\eta}, \frac{d\phi}{dr} = \phi_0 e^{2\omega_0}\frac{\left(1-\frac{\eta}{r}\right)^{-4m/\eta}}{r^2}, \quad (28)$$

$$\sigma^2 = \frac{4m^2-\eta^2}{2\lambda}, \lambda = \tfrac{1}{2}(4\xi-3). \quad (29)$$

The solution (27) looks pretty similar to eq.(23), but not quite. The horizon appears at r=η. The curvature in the static frame is finite, but in the moving frame, we find

$$\left|\frac{R''}{R}\right| = \frac{m(m-\eta)}{r^4\left(1-\dfrac{\eta}{r}\right)^2}. \tag{30}$$

Therefore, $R^{(ex)}_{0202} \to \infty$ as $r \to \eta$, unless $m = \eta$. Let us examine what happens to the Weyl scalar R. It is given by

$$R = -r^{-\left(2+\frac{2m}{\eta}\right)}(r-\eta)^{-3+\frac{2m}{\eta}} p(r), \tag{31}$$

where p(r) is a polynomial of O(r )>0. If this scalar is finite, so is the Riemann scalar. If $m=\eta$, then $R \to \infty$ as $r \to \eta$. For $m \neq \eta$, tidal forces in the freely falling frame becomes infinitely large. In either case, the horizon is singular.

The two types of solutions (23) and (27) following from two entirely different theories exhibit a remarkable similarity as far as the behavior of tidal forces are concerned. In the examples considered so far, it is clear that the black hole solutions of EGR (with or without source) are indeed black holes while those from the non-EGR are different, at least as far as tidal force considerations are concerned. Can we generalize our conclusion to include all SSS solutions of EGR? Perhaps not. For instance, consider the Janis-Newman-Winnicour [12] solution of the Einstein minimally coupled theory. It has exactly the same form as that of eq.(23) with only a different (logarithmic) scalar field. Consequently, the tidal forces in the geodesic frame are infinite near the horizon. That explains why the JNW solution is said to have a naked singularity [13], but it can also be interpreted as having the features of a CBH. At any rate, it follows that every EGR solution should be tested on a case by case basis.

## 5. Kruskal-like extension of CBH

It is well known that Kruskal-Szekeres extension [14,15] offers the advantage that it removes the coordinate singularity from the Schwarzschild metric in the standard form. Let us now ask if similar advantages obtain in the Kruskal-like extension of the metric (20) performed by Campanelli and Lousto [6]. The answer seems to be in the negative. Defining the null variables $\bar{u}, \bar{v}$ by [16]

$$d\bar{u} = dt - dr^*, \quad d\bar{v} = dt + dr^*, \tag{32}$$

where the tortoise-like variable $r^*$ is given by

$$dr^* = \left(1-\frac{2m}{r}\right)^{\frac{\beta-\alpha}{2}-1} dr, \tag{33}$$

and applying further the transformations

$$U = -\exp(-\kappa_H)\bar{u}, \quad V = \exp(\kappa_H \bar{v}), \tag{34}$$

in which the surface gravity $\kappa$ is given by

$$\kappa = (\alpha+1)\frac{m}{r^2}\left(1-\frac{2m}{r}\right)^{\frac{\alpha-\beta}{2}}, \qquad (35)$$

the final form of the metric (20) becomes [17]:

$$ds^2 = -\left(1-\frac{2m}{r}\right)^{\alpha+1}\kappa_H^{-2}\exp(-2\kappa_H r^*)dUdV + r^2\left(1-\frac{2m}{r}\right)^{\beta}d\Omega^2, \qquad (36)$$

where $\kappa_H$ is the value of $\kappa$ at r=2m. Thus, $\kappa_H = 0$ for $\alpha>\beta$, $\kappa_H = \infty$ for $\alpha<\beta$, $\kappa_H = (\alpha+1)/4m$ for $\alpha=\beta$. One may apply a further transformation to spacelike and timelike variables u = (V-U)/2, v = (V+U)/2 respectively on metric (36) to make it look partly familiar. For the special case $\alpha=\beta$, one has only $g_{UV}(r=2m)$ finite. For $\alpha\neq\beta$, which includes $\alpha=-\beta$ [leading to the metric (23)], $g_{UV}(r=2m)$ is no longer finite. But most importantly, the last term in the metric (36) continue to have a singularity at r=2m for $\beta\leq-1$. Thus, the Kruskal-like extension, eqn. (36), does *not* offer any advantage as such. The surface area of the horizon still remains infinite. One may then equally well use metrics (20) and (23) for computing curvature tensors and invariants, which was actually done in Sec. 4e. Similar considerations apply to the metric (27).

The metric representing type B1 CBHs proposed by Bronnikov et al [7] is conformal to the metric (20). As shown in Ref.[7], the tidal forces are infinite at the horizon in this case, too. The surface area of the horizon remains infinite even in the extended form of the metric. It is not possible to reduce the area by Kruskal-like extensions.

## 6. A better CBH

We have seen that Kruskal-like extensions can not do away with the singularities in $g_{\theta\theta}$ and $g_{\varphi\varphi}$ as is evident from the metric (36). Consequently, one has infinite horizon areas and entropies. Hence the name CBH, as mentioned before. Accepting these facts, we enquire if there exists a CBH for which the tidal forces are *finite* on the horizon. It seems that there indeed is one: The class IV solutions of the BD theory [18] provide just such a CBH.

Class IV solutions are:

$$ds^2 = -e^{2\mu(r)}dt^2 + e^{2\nu(r)}\left[dr^2 + r^2d\Omega^2\right], \qquad (37)$$

$$\mu(r) = \mu_0 - \frac{1}{Br}, \qquad (38)$$

$$\nu(r) = \nu_0 + \frac{C+1}{Br}, \qquad (39)$$

$$C = \frac{-1\pm\sqrt{-2\varpi-3}}{\varpi+2}, \qquad (40)$$

$$\phi = \phi_0 e^{-\frac{C}{Br}}. \tag{41}$$

Usual asymptotic flatness and weak field conditions fix the constants as

$$\mu_0 = v_0 = 0, B = 1/m > 0, \tag{42}$$

where m is the mass of the configuration. The horizon appears at r=r$_h$=0 and its area is infinite. Other features of this solution are discussed in Refs.[19,20]. In the static orthonormal frame, the curvature components are:

$$R_{0101} = -\frac{1}{Br^3 e^{\frac{2(C+1)}{Br}}} \left[ \frac{C+2}{Br} - 2 \right], \tag{43}$$

$$R_{0202} = R_{0303} = -\frac{1}{Br^3 e^{\frac{2(C+1)}{Br}}} \left[ 1 - \frac{C+1}{Br} \right], \tag{44}$$

$$R_{1212} = R_{1313} = -\frac{C+1}{Br^3 e^{\frac{2(C+1)}{Br}}}, \tag{45}$$

$$R_{2323} = -\frac{C+1}{Br^3 e^{\frac{2(C+1)}{Br}}} \left[ \frac{C+1}{Br} - 2 \right]. \tag{46}$$

All these components tend to zero as r→r$_h$, provided C+1≥0. This happens only if ϖ<-2. It should be recalled that the CBH solutions proposed in Refs. [6] and [7] also correspond to negative values of ϖ. As discussed in Ref.[6], it is the numerical value of ϖ, rather than its sign, that is more relevant. Also, the EGR effects are recovered for $|\varpi| \to \infty$.

The Ricci scalar for the considered solution is given by

$$R \equiv g_{\alpha\beta} R^{\alpha\beta} = \frac{-2(1+C+C^2)}{B^2 r^4} e^{-\frac{2(C+1)}{Br}} \tag{47}$$

which goes to zero as r→r$_h$. It may be verified that all other Riemann invariants are also finite at the horizon. The metric (37)-(41) has a remarkable feature in that, for C=0, it describes all the solar system tests exactly as does the EGR Schwarzschild metric. The curvature components in the moving frame also remain finite as the horizon is approached. For example,

$$R_{\hat{0}\hat{2}\hat{0}\hat{2}} = \left( \frac{C+1}{Br^2} - \frac{1}{r} \right) \left( \frac{e^{-\frac{2C}{Br}}}{Br^2} \right) \left( CE_s^2 - (C+1)e^{-\frac{2}{Br}} \right)$$

$$-\left(\frac{e^{-\frac{2C}{Br}}}{Br^2}\right)\left(\frac{C+1}{Br^2}+\frac{C}{r}\right)E_{ex}^2, \tag{48}$$

tends to zero as r→$r_h$. This implies that the tidal forces in the geodesic frame do not diverge. In many ways, therefore, the class IV metric resembles the ones discussed in Refs. [6-8] but it has an added merit as indicated above. Hence, it seems to represent a better CBH than the ones proposed so far.

Summarizing, we have to say the following: We examined a few EGR solutions for which the tidal forces do not diverge on the horizon. However, it was also indicated that a general conclusion to that effect can not be drawn because of the existence of the JNW solution. Non-EGR CBH solutions exhibit infinite tidal forces on the horizon. Extended solutions too fail to remove this divergence as the CBH metric continues to remain singular. We then presented a better CBH in the BD theory. Bronnikov et al [7] conjectured that infinite horizon areas could be related to infinite tidal forces. Our example in Sec.6 indicates that this is not necessarily the case. In that sense, class IV solutions may be interpreted as providing a counterexample to the conjecture.

**Acknowledgments**

We are grateful to Professor Gary T. Horowitz for an inspiring correspondence. One of us (KKN) wishes to thank Inter-University Center for Astronomy and Astrophysics (IUCAA), Pune, India, where part of the work was carried out.